\documentstyle[eqsecnum,prd,aps,floats,twocolumn,epsfig]{revtex}

\newcommand{\beq}{\begin{equation}}
\newcommand{\eeq}{\end{equation}}
\newcommand{\bea}{\begin{eqnarray}}
\newcommand{\eea}{\end{eqnarray}}

\def \pa {\partial}
\def \ra {\rightarrow}
\def \la {\lambda}

\def \Da {\Delta}
\def \b {\beta}
\def \a {\alpha}
\def \ap {\alpha^{\prime}}
\def \Ga {\Gamma}

\def \sg {\sigma}
\def \da {\delta}

\def \bp {\dot{\beta}}
\def \bpp {\ddot{\beta}}
\def \fp {\dot{\phi}}
\def \fpp {\ddot{\phi}}
\def \hp {\dot{h}}
\def \hpp {\ddot{h}}

\begin{document}

\par
\begingroup
\twocolumn[%

\begin{flushright}
DFTT-23/97\\
gr-qc/9704045
\end{flushright}

{\large\bf\centering\ignorespaces
Tensor perturbations in high-curvature string backgrounds
\vskip2.5pt}
{\dimen0=-\prevdepth \advance\dimen0 by23pt
\nointerlineskip \rm\centering
\vrule height\dimen0 width0pt\relax\ignorespaces
M. Gasperini
\par}
{\small\it\centering\ignorespaces
Dipartimento di Fisica Teorica, Universit\`a di Torino, 
Via P. Giuria 1, 10125 Turin, Italy\\
and Istituto Nazionale di Fisica Nucleare, Sezione di Torino, 
Turin, Italy\\ 
\par}
{\small\rm\centering(\ignorespaces  April 1997\unskip)\par}

\par
\bgroup
\leftskip=0.10753\textwidth \rightskip\leftskip
\dimen0=-\prevdepth \advance\dimen0 by17.5pt \nointerlineskip
\small\vrule width 0pt height\dimen0 \relax

We derive a generalized equation for the evolution of tensor 
perturbations in a cosmological background, taking into account 
higher-curvature contributions and a tree-level coupling to the dilaton 
in the string frame. The equation is obtained by perturbing the 
gravi-dilaton string effective action, expanded up to first order in 
$\alpha'$. The $\alpha'$ corrections can modify the low-energy 
perturbation spectrum, but the modifications are shown to be small when 
the background curvature keeps constant in the string frame.

\par\egroup
\vskip2pc]
\thispagestyle{plain}
\endgroup

\section{INTRODUCTION}
\label{I}

The generation of a primordial perturbation spectrum, able to explain 
the observed large scale anisotropy, is one of the most celebrated 
aspects of the present inflationary cosmological models \cite{1}.

Such a spectrum is obtained through the amplification of the quantum 
fluctuations of the metric and of the background sources around their 
initial configuration \cite{2,2a}, and is usually computed in the context 
of a lowest-order, scalar-tensor effective theory of gravity. 
Higher-derivative (i.e. higher-curvature) corrections are usually 
neglected or, when included, are parametrized by arbitrary functions of 
the scalar curvature \cite{2,3} (see \cite{4} for a possible 
exception),  without consequence for the primordial spectrum
since the  effective perturbation equations are left unchanged. 

The absence of higher-curvature contributions, however, 
is hard to swallow for a perturbation spectrum that originates just in 
the primordial cosmological phase when the Universe, according to the 
standard scenario, is highly curved and is expected to approach the 
Planck scale and the quantum gravity regime. It would seem more 
appropriate, in such a context, to include all possible high-curvature 
corrections into the effective action, and ask whether their inclusion 
can in general modify the primordial spectrum obtained from the 
lowest-order perturbation equations. 

The answer to this question is, unfortunately, model-dependent, being 
subordinate to the explicit form of the higher-order terms added to the 
action. The aim of this paper is to discuss higher-curvature corrections 
to the tensor perturbation spectrum, in the particular case of the 
gravi-dilaton effective action of string theory \cite{5}. In the 
so-called ``pre-big bang" cosmological models \cite{6} obtained in that 
context, the occurrence of a high-curvature string phase in which 
higher-derivative corrections cannot be neglected is indeed an 
unavoidable consequence of the initial conditions, chosen to approach 
the string perturbative vacuum \cite{7}. Fortunately, in that case, the 
corrections cannot be added arbitrarily, but are rigidly prescribed by 
the ``$\ap$-expansion" of the string effective action \cite{8}. 

We shall limit, in this paper, to the first order in $\ap$, namely to 
the four-derivative terms corresponding to quadratic curvature 
corrections. Already at this level, as we will show, the amplification 
of the tensor fluctuations of the vacuum is governed by a modified 
perturbation equation. In a background with constant curvature and 
linearly evolving dilaton, like in the string phase 
\cite{9}--\cite{11} typical of the pre-big bang scenario , the
differences between the  corrected higher-order equation and
the approximated low-energy equation  are small, however,
and can be neglected for an order-of-magnitude  estimate of
the graviton spectrum. In other backgrounds, with running 
curvature and long enough duration of the high-curvature
phase,  the modified equation found in this paper may lead
instead  to important differences in the perturbation spectrum.

The paper is organized as follows. In Sec. \ref{II} we present the 
background equations and we perturb the gravi-dilaton effective action, 
up to second order in the tensor perturbation variable, including the 
quadratic curvature terms prescribed by string theory. The 
diagonalization of the perturbed action defines a new canonical 
variable, for the normalization of the spectrum to the quantum 
fluctuations of the vacuum. In Sec. \ref{III} we discuss the 
time-evolution of the normalized canonical variable in a high-curvature 
string background, and we estimate the resulting tensor perturbation 
spectrum. Sect. \ref{IV} contains a brief summary and our
concluding  remarks. The details of the perturbative
computation are presented in Appendix \ref{A}.

\section{BACKGROUND AND PERTURBATION EQUATIONS}
\label{II}

In the string frame, and to first order in the high-derivative 
$\ap$ expansion, the effective action that reproduces the gravi-dilaton 
sector of the tree-level string $S$-matrix can be written in the form 
\cite{8}:
\bea
&&S=\int d^{4}x \sqrt{-g}e^{-\phi} \nonumber\\
&& \times\left\{- R-
\pa_\mu \phi\pa^\mu \phi+{k\ap \over 4} \left[R^2_{GB} - 
(\pa_\mu\phi\pa^\mu\phi)^2\right]\right\}.
\label{21}
\eea
Here $\phi$ is the dilaton field, $\ap=\la_s^2$ is the fundamental 
string-length parameter governing the importance of the higher-curvature 
corrections, $k=1,2$ for the bosonic and heterotic string, respectively, 
and  $R^2_{GB} 
\equiv R_{\mu\nu\a\b}^2-4  R_{\mu\nu}^2+
R^2$ is the usual Gauss-Bonnet invariant (conventions: 
$g_{\mu\nu}=(+---)$, $R_{\nu\a}=R_{\mu\nu\a}\,^\mu$, and 
$R_{\mu\nu\a}\,^\b=\pa_\mu\Ga_{\nu\a}\,^\b-...$). 
Note that we have chosen a field redefinition that removes terms with 
higher-than-second derivatives from the effective equations \cite{8,12}. 
Also, we are considering the case of critical (super)strings, in which 
no effective cosmological term appears in the action. This means, in 
$d=3$ spatial dimensions, 
that some ``passive" sector is assumed to be present to cancel the 
central charge deficit $(3-d_{{\rm crit}})/\ap$. All the computations of 
this paper can be easily extended, however, to include a non-vanishing 
cosmological constant.

We shall consider perturbations around a $d=3$, spatially
flat background,  parametrized by
\beq
ds^2=N^2(t)dt^2-a^2(t)dx_i^2 , ~~ \phi=\phi(t),~~ a(t)=e^{\b(t)}
\label{22}
\eeq
($i,j=1,2,3$). For such a background the action (\ref{21}) becomes, after
integration by parts (see Appendix \ref{A}):
\bea
&& S=\int dt e^{3\b-\phi}\Bigg[{1\over N} \left(6\bp\fp-6\bp^2-
\fp^2\right)\nonumber\\
&& \qquad 
+ {\ap\over 4 N^3}\left(8\fp\bp^3 -\fp^4 \right)\Bigg].
\label{23}
\eea
A dot denotes differentiation with respect to $t$, and we have
put $k=1$, for simplicity. By varying the action with respect to $N$,
$\b$ and $\phi$, and imposing the cosmic time gauge $N=1$, we get
respectively the equations:
\beq
6H^2+\fp^2-6H\fp-{3\over 4}\ap (8\fp H^3-\fp^4)=0,
\label{24}
\eeq
\bea
&&
-18H^2-3\fp^2+12H\fp+6\fpp-12\dot H \nonumber\\
&&
+\ap (12\fp H^3+{3\over 4}\fp^4-6H^2\fp^2+6H^2\fpp+
12\fp H \dot H)
=0,
\label{25}
\eea
\bea
&&
12H^2+\fp^2-6H\fp-2\fpp+6\dot H \nonumber\\
&&
+\ap (6 H^4+{3\over 4}\fp^4-3H\fp^3+6H^2 \dot H-3\fp^2 \fpp )=0
\label{26}
\eea
(we have adopted the usual notation for the Hubble parameter, $H=\dot
a/a\equiv \bp$, $\dot H = \bpp$). Their solutions provide the
gravi-dilaton background for the propagation of metric perturbations.
Note that only two of these equations are independent \cite{11}
(unless $\fp=3\bp$), and that the first equation, following from the
variation of the lapse function, can be used as  a ``non-dynamical"
constraint on the set of initial conditions. 

In this paper we shall restrict our attention to tensor metric
perturbations, parametrized by the transverse, trace-free variable
$h_{\mu\nu}$,
\beq
g_{\mu\nu} \ra  g_{\mu\nu}+ \da g_{\mu\nu},~\,
\da g_{\mu\nu} = h_{\mu\nu}, ~ \,
\nabla_\nu h_\mu\,^\nu =0=h_\mu\,^\nu,
\label{27}
\eeq
where $\nabla _\mu$ denotes covariant differentiation with respect to
$g_{\mu\nu}$, and the indices of $ h_{\mu\nu}$ are raised and lowered
with the unperturbed metric, $h_\mu\,^\nu=g^{\nu\a}h_{\mu\a}$. By
expanding, up to second order in $h$, the controvariant components of
the metric tensor,
\beq
\da^{(1)} g^{\mu\nu}=-h^{\mu\nu}, ~~~~ \da^{(2)}
g^{\mu\nu}=h^{\mu\a}h_\a\,^\nu 
\label{28}
\eeq
($\da^{(k)}A$ denotes the $k$-th term in the expansion of a variable
$A$ in powers of $h$), of the volume density,
\beq
\da^{(1)} \sqrt{-g}=0, ~~~~ \da^{(2)} \sqrt{-g}=-{1\over
4}\sqrt{-g}h_{\mu\nu}h^{\mu\nu},  
\label{29}
\eeq
and of the components of the Riemann and Ricci tensor (see Appendix
\ref{A}), we will obtain from (\ref{21}) an action quadratic in the
perturbation variable $h_\mu\,^\nu$, governing the dynamic of tensor
perturbations in the corresponding gravi-dilaton background. The
method is exactly the same as the one first used in \cite{13} for
determining the effective action of tensor perturbations in a
cosmological background (see also \cite{2,14}), with the only
difference that in the present case the perturbed action includes
higher-curvature terms, and a non-minimal coupling of the metric to
the dilaton field. 

It is convenient to work in the synchronous gauge, where
\bea
&& 
g_{00}=1, ~~~g_{0i}=0, ~~~ g_{ij}=-a^2\da_{ij}, \nonumber \\
&&
h_{00}=0, ~~~h_{0i}=0, ~~~ g^{ij}h_{ij}=0, ~~~ \pa_j h_i\,^j =0 .
\label{210}
\eea
After repeated use of the identities (see Appendix \ref{A})
\bea
&&
g^{jk}\hp_{ik}=\hp_i\,^j +2 H h_i\,^j , \nonumber\\
&&
g^{jk}\hpp_{ik}=\hpp_i\,^j +2 \dot H h_i\,^j +4H\hp_i\,^j +4H^2h_i\,^j, 
\label{211}
\eea
we can express the second-order variation of the action (\ref{21}) as a
quadratic form depending on the first and second derivatives of the
symmetric, trace-free matrix $h\equiv h_i\,^j$, with time-dependent
coefficients fixed by the background fields $a(t)$, $\phi (t)$:
\bea
&&\qquad
\da^{(2)} S\equiv \nonumber\\
&& \qquad
\int d^4x e^{-\phi} \Bigg[
-\da^{(2)}\left(\sqrt{-g} R\right)-
\da^{(2)}\left(\sqrt{-g}
\pa_\mu\phi\pa^\mu\phi\right)\nonumber\\ &&\qquad
+{\ap \over 4} \da^{(2)}\left(\sqrt{-g} R_{GB}^2\right)-
{\ap \over 4} 
\da^{(2)}\left(\sqrt{-g} \left(\pa_\mu\phi
\pa^\mu\phi\right)^2\right)
\Bigg]\nonumber\\
&&\qquad
=\int d^4x e^{-\phi} a^3 {\rm Tr}~ \Bigg[h^2({1\over 4}\fp^2-
{3\over 2}\dot H - 3 H^2)\nonumber\\
&&\qquad
-h\hpp-4Hh\hp-{3\over 4} \hp^2+{1\over 4}h {\nabla^2 \over a^2}h 
\Bigg]\nonumber \\
&&\qquad
+{\ap \over 4}\int d^4x e^{-\phi} a^3 {\rm Tr}~ \Bigg[h^2({1\over
4}\fp^4- 6\dot H H^2- 6 H^4)\nonumber\\
&&\qquad
-4 H^2 h\hpp -h\hp (8H\dot H+16H^3)-\hp^2(\dot H +7H^2)
\nonumber\\
&&\qquad
-2H\hp\hpp
 +(2\hpp+\dot H h +4H\hp+H^2h)
{\nabla^2 \over a^2}h \nonumber\\
&&\qquad
+2\hp {\nabla^2 \over a^2}\hp\Bigg] . 
\label{212}
\eea
Here $\nabla^2=\da^{ij}\pa_i\pa_j$ is the flat-space Laplace
operator,  ${\rm Tr}~h^2\equiv h_i\,^jh_j\,^i$, and so on (the 
explicit computation of the various terms is reported in  Appendix
\ref{A}). Integrating by parts the terms with more than two partial
derivatives acting on $h$, as well as the terms in $h\hp$ and $h\hpp$,
we can put the action in the convenient form
\bea
&&\qquad
\da^{(2)} S\equiv 
\int d^4x e^{-\phi} a^3 {\rm Tr}~\Bigg\{{1\over 4}\hp^2(1-\ap H \fp) 
\nonumber\\
&&\qquad
+{1\over 4} h{\nabla^2 \over a^2}h \left[1+\ap(\fp^2-\fpp)\right]+
h^2\Bigg[ {1\over 2}\fpp+H\fp 
\nonumber\\
&&\qquad
-{1\over 4}\fp^2 -\dot H -{3\over 2} H^2 +{\ap\over 4}\Bigg(
{1\over 4} \fp^4+2H^2\fpp
\nonumber\\
&&\qquad
-2H^2\fp^2+4\fp \dot H H + 4 \fp H^3 \Bigg)\Bigg]\Bigg\}.
\label{213}
\eea
The absence of terms with more than two derivatives follows from the
fact that the higher-curvature corrections appear in the action as an
Euler form (the Gauss-Bonnet invariant). Note also that all
$\ap$ corrections disappear in the limit $\phi$=const, since in
that case the higher-curvature part of the action (\ref{21})
reduces (in $d=3$) to a total derivative that does not
contribute to the variation. 

The coefficient of the $h^2$ term in eq. (\ref{213}) is identically
vanishing, thanks to the background equation (\ref{25}). By
decomposing the matrix $h_i\,^j$ into the two physical polarization
modes of tensor perturbations, $h_+$ and $h_\times$, 
\beq
{\rm Tr}~h^2\equiv h_i\,^jh_j\,^i=
2\left(h_+^2+h_\times^2\right), \label{214}
\eeq
we can finally write the action, for each polarization mode
$h(x,t)$, as 
\bea
&&\qquad
\da^{(2)} S_h= {1\over 2} 
\int d^4x e^{-\phi} a^3\Bigg\{ \hp^2\left(1-\ap H\fp\right) +
\nonumber\\
&&\qquad
+ h{\nabla^2 \over a^2}h\left[1+\ap\left(\fp^2-\fpp\right)\right]
\Bigg\}, 
\label{215}
\eea
where $h$ is now a scalar variable standing for either one of the two
polarization amplitudes $h_+$, $h_\times$. The variation of the action
with respect to $h$ gives then the modified perturbation equation:
\bea
&&\qquad
\hpp\left((1-\ap H\fp\right)+\hp \Bigg[
3H-\fp-\ap\Bigg(3H^2\fp-H\fp^2
 \nonumber\\
&&\qquad
+\dot H \fp
+H\fpp\Bigg)\Bigg] -{\nabla^2 \over a^2}h \left[1+\ap
\left(\fp^2-\fpp\right)\right]=0.
\label{216}
\eea
In the absence of $\ap$ corrections, and for a constant dilaton
background, we recover the well-known result $\Box h=0$, describing
the propagation of a massless scalar degrees of freedom minimally
coupled to the background metric \cite{2}, \cite{13}-\cite{15}. When
$\ap=0$, and $\fp\not=0$, we recover instead the perturbation
equation in a Brans-Dicke background \cite{16},  $(\Box-\fp)h=0$,
describing the propagation of gravity waves in the string frame
according to the lowest-order string effective action. 

Eq. (\ref{216}) controls the time evolution of the Fourier
components $h_k$ of the two polarization modes. In order to
normalize the spectrum to the quantum fluctuations of the
vacuum, however, we need the canonical variable that
diagonalizes the perturbed action \cite{2,17}, and that
represents in this case the normal modes of tensor oscillations
of our gravi-dilaton background. We note, to this purpose, that
introducing the conformal time coordinate $\eta$, defined by
$a=dt/d\eta$, the action (\ref{215}) can be written in the form
\beq \da^{(2)} S_h= {1\over 2} 
\int d^3x d\eta \left[z^2(\eta)h^{\prime 2}+ y^2(\eta)h\nabla^2
h\right],
\label{217}
\eeq
where a prime denotes differentiation with 
respect to $\eta$, and
\bea
&&
z^2(\eta)=e^{-\phi}\left(a^2-\ap {a'\over a}\phi'\right),
\nonumber\\
&&
y^2(\eta)=e^{-\phi}\left[a^2+\ap 
\left(\phi^{\prime 2}-\phi^{''}+
 {a'\over a}\phi'\right)\right].
\label{218}
\eea
By setting $\psi=zh$ the action becomes
\beq
\da^{(2)} S_h= {1\over 2} 
\int d^3x d\eta \left(\psi^{\prime 2}+ {z^{''}\over z} \psi^2+
{y^2\over z^2}\psi\nabla^2 \psi \right). 
\label{219}
\eeq
For each Fourier mode we can thus define a canonical variable, 
$\psi_k=zh_k$, that diagonalizes the kinetic part of the action, and
that satisfies an evolution equation of the usual form \cite{2,15},
\bea
&&
\psi_k^{\prime\prime}+\left[k^2-V_k(\eta)\right]\psi_k=0,
\nonumber \\ &&
V_k(\eta)={z^{''}\over z}-{k^2\over z^2}(y^2-z^2),
\label{220}
\eea
with the only difference that the effective potential 
$V_k(\eta)$ is, in general, $k$-dependent. 

This equation, that encodes into the effective potential the
higher-curvature corrections (through eq. (\ref{218})), 
represents the
main result of this paper, and will be used in Sec. \ref{III} to
discuss the amplification of tensor fluctuations in the gravi-dilaton
background of a typical string cosmology model. 

\section{HIGHER-CURVATURE CONTRIBUTIONS TO THE GRAVITON 
SPECTRUM}
\label{III}

In the context of the pre-big bang scenario \cite{6,7}, typical of
string cosmology, the background equations
(\ref{24})--(\ref{26}) describe the evolution of the Universe from
an asymptotic initial state with $H=0$ and $\fp=0$ (the string
perturbative vacuum). 

As long as the space-time curvature and the dilaton kinetic
energy are small in string units, $\ap H^2\ll 1$, $\ap \fp^2\ll 1$,
the higher-order $\ap$ corrections can be neglected, and the
low- energy solutions \cite{18} of the background equations
describe an accelerated growth of the curvature and of the
string coupling, with $H>0$, $\dot H>0$, $\fp>0$. As soon as the
curvature reaches the string scale, however, the effect of the
$\ap$ corrections tends to stabilize the background in a phase
of constant curvature and linearly evolving dilaton, $H=$const,
$\fp=$const, as recently discussed in \cite{11} (the final
transition to the standard, decelerated evolution eventually
occurs when the radiation back-reaction becomes important,
and generates quantum loop corrections to the effective action
\cite{18a}). The typical accelerated evolution of $H$ and $\fp$,
obtained through a numerical integration of eqs. 
(\ref{24})--(\ref{26}) with the perturbative vacuum as initial
condition at $t \ra -\infty$, is shown in Fig. 1. 

\begin{figure}[t]
\begin{center}
\mbox{\epsfig{file=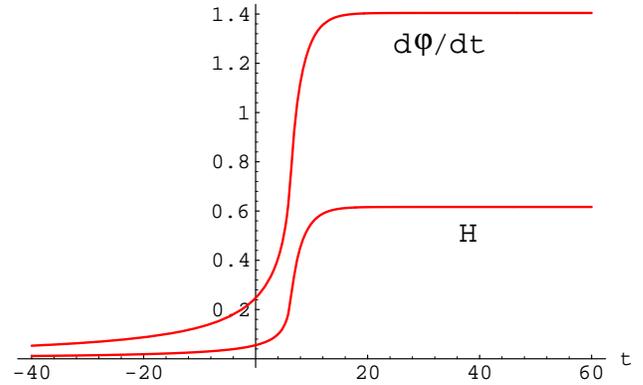,width=82mm}}
\vskip 5mm
\caption{Time evolution of the gravi-dilaton background, with 
the perturbative vacuum as initial condition at $t \ra -\infty$ (in
units $\ap=1$). The plot shows the results of a numerical
integration of the string cosmology equations
 (\ref{24})--(\ref{26}).}
\end{center}
\end{figure}

As clearly shown in the figure, the inflationary evolution of this
class of backgrounds can be sharply divided into two distinct
regimes: an initial dilaton-driven phase, in which the $\ap$
corrections are negligible, and a high curvature string phase, in
which the $\ap$ corrections are dominant and stabilize the
background curvature at the string scale. In the first phase eq.
(\ref{220}) reduces to the usual perturbation equation (including
a tree-level coupling to the dilaton \cite{16}):
\beq
\psi_k^{\prime\prime}+\left(k^2-{\xi^{''}\over
\xi}\right)\psi_k=0,\,\,\, \,\,\xi=a e^{-\phi/2}.
\label{31}
\eeq
In the second phase the $\ap$ corrections cannot be neglected,
but the background curvature is constant,
$H=a'/a^2=c_1/\sqrt{\ap}$ and $\fp=\phi'/a=c_2/\sqrt{\ap}$, so
that 
\beq
z^2=\xi^2(1-c_1c_2), \,\,\,\,\,\,\,\, 
y^2=\xi^2(1+c_2^2).
\label{32}
\eeq
The only change of eq. (\ref{31}) is thus an effective shift of the
comoving frequency:
\bea
&&
\psi_k^{\prime\prime}+\left[k^2(1+c)-{\xi^{''}\over
\xi}\right]\psi_k=0,\,\,\, \,\,\xi=a e^{-\phi/2}, \nonumber\\
&&
c= \ap{H\fp+\fp^2\over 1-\ap H\fp}={c_1c_2+c_2^2\over
1-c_1c_2}= {\rm const}.
 \label{33}
\eea

The important consequence of the above equations is that, for
the whole class of background that we are considering, 
no modification is induced by the high-curvature terms on the 
evolution of $\psi_k$ outside the horizon (i.e. for
$|k\eta|\ll1$). Such an evolution is uniquely determined, both in
the low- and high-curvature regime, by the time-behaviour of
the background variable $\xi(\eta)$, according to the
asymptotic solutions of eqs. (\ref{31}) and (\ref{33}):
\beq
\psi_k= A_k\xi(\eta)+B_k\xi(\eta)\int^\eta d\eta'\xi^{-2}(\eta') 
\label{34}
\eeq
($A_k,B_k$ are integration constants). Hence,  no
modification is induced in the perturbation
spectrum (both for modes crossing the horizon in the
dilaton-driven and in the string phase),   
compared with the spectrum determined without 
the $\ap$ corrections in the perturbation equation. 

Let us consider, in particular, the string phase, assuming that
$\psi (\eta)\sim \xi(\eta)\sim (-\eta)^\a$, $\a\leq 1/2$, is the
dominant term in the asymptotic solution (\ref{34}) for $\eta
\ra 0_-$. By using the correct
normalization of the canonical variable \cite{2} at horizon
crossing ($hc$), $|\psi_k|_{hc}=1/\sqrt k$, we find indeed, for
$\eta \ra 0_-$, the power spectrum
\beq
k^{3/2} |\psi_k|=k{\xi\over \xi_{hc}}=k|k\eta\sqrt{1+c}|^\a \sim
k^{1+\a}, 
\label{35}
\eeq
which is exactly the same as that provided by the low-energy
perturbation equation. The only effect of the high-curvature
terms is the shift $\Da \psi_k/\psi_k$ of the asymptotic
amplitude, due to a shift of the horizon crossing scale,
\beq
{|\Da \psi_k|\over |\psi_k|}= \left|(\sqrt{1+c})^\a -1\right|.
\label{36}
\eeq
Such a shift is the same for all modes, and is of order one if, as
expected, the constant values of $H$ and $\fp$ are of order one
in string units (unless $c_1$ and $c_2$ are both exactly equal to
one, in which case $c\ra \infty$, see eq. (\ref{33})).

The above conclusions are confirmed by a numerical integration
of the system formed by the background equations
(\ref{24})--(\ref{26}) and by the exact perturbation equation
(\ref{216}). The results are shown in Fig. 2 where we have
compared, for a mode crossing the horizon in the string phase,
the evolution in cosmic time of the amplitude $|h_k(t)|$
obtained from the exact perturbation equation (\ref{216}),  with
the amplitude that one would obtain (for the {\it same} mode
$k$ and in the {\it same} background) neglecting the $\ap$
corrections in the perturbation equation. In both cases the
amplitude oscillates outside the horizon, and the oscillations are
damped outside the horizon, as expected. The effect of the $\ap$
corrections, when they become important, is to induce an
effective shift of the comoving frequency, with a resulting shift
of the final asymptotic amplitude, as clearly shown by the
figure. Since the shift is typically of order one, the previous
computations \cite{10} concerning graviton production during
the string phase (see also \cite{7,19,19a}), performed without
$\ap$ corrections, remain valid as an order of magnitude
estimate.

\begin{figure}[t]
\begin{center}
\mbox{\epsfig{file=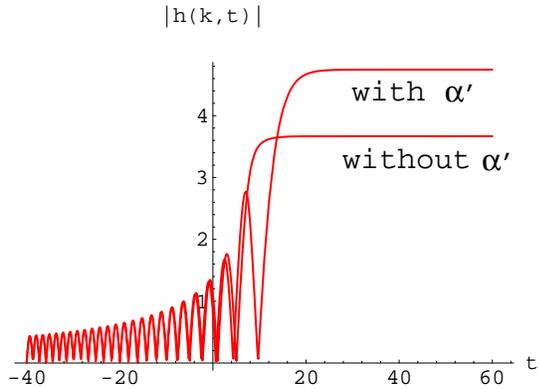,width=82mm}}
\vskip 5mm
\caption{Time evolution of $|h_k(t)|$ for a mode of comoving
frequency $k=1$ (in units $\ap=1$), in the background solution
corresponding to Fig. 1. The evolution is obtained through a
numerical integration of the perturbation equation (\ref{216}),
with and without $\ap$ corrections.}
\end{center}
\end{figure}

It is important to stress that this conclusion is valid provided
$H$ and $\fp$ stay constant for the whole duration of the
high-curvature string phase. In the opposite case the $\ap$
corrections may affect the time-evolution outside the horizon
and, as a consequence, the perturbation spectrum. Of course, in
a general background in which the growth of the curvature
scale is unbounded, our modified perturbation equation can
only be applied for $\ap H\fp<1$ (at higher scales, higher
powers of the curvature should be added). The importance of
the $\ap$ corrections can be consistently checked, however,
even inside the range of validity of our equations, as
illustrated in Fig. 3. 

The figure shows the result of a numerical integration of eq.
(\ref{216}), with and without $\ap$ corrections, for the typical
power-law background $a=(-t)^{-1/2}$, $\phi= -3\ln (-t)$ (which
corresponds, in the low-energy limit, to a background
dominated by a perfect gas of unstable, stretched strings
\cite{6,18}, with equation of state $p=-\rho/3$). In the absence
of $\ap$ corrections the solution of eq. (\ref{216}) gives an
amplitude $h_k(t)$ which grows, asymptotically, like
$(-t)^{-1/2}$. With the $\ap$ terms included, the asymptotic
behaviour of the amplitude is significantly different, and the
differences are $k$-dependent, leading to a final modified
spectrum. Note that the induced shift  
is amplified in time when the comoving amplitude is not
asymptotically constant, like for the case illustrated in Fig. 3. 

\begin{figure}[t]
\begin{center}
\mbox{\epsfig{file=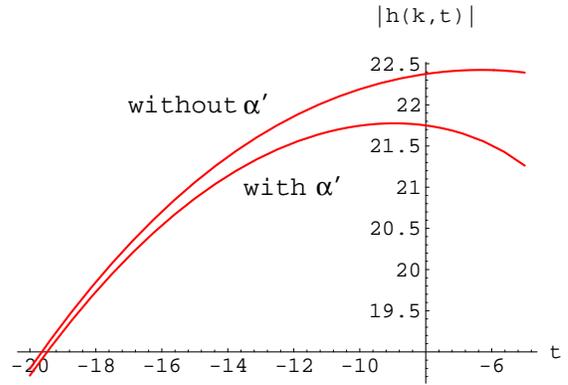,width=82mm}}
\vskip 5mm
\caption{Time evolution of $|h_k(t)|$ outside the horizon, for a
mode of comoving frequency $k^2=0.005$ (in units $\ap=1$). The
evolution is obtained through a numerical integration of 
eq. (\ref{216}), with and without $\ap$ corrections, in the
background $a=(-t)^{-1/2}$, $\phi= -3\ln (-t)$.}
\end{center}
\end{figure}

We come back, finally, to the case in which the high-curvature
phase is frozen at the string scale, and the corrected
perturbation equation takes the form (\ref{33}). The
perturbation spectrum, for modes crossing the horizon in the
string phase, is  determined by the corresponding
evolution of $\xi(\eta)$. By using the asymptotic values of $H$
and $\fp$ given by the model of background discussed in this
paper,
\beq
c_1=\sqrt{\ap} H=0.616...., \,\,\,\,\,\,
c_2=\sqrt{\ap} \fp=1.40....
\label{37}
\eeq
(see Fig. 1, and also \cite{11}), we find
\bea
&&
a(\eta)= -{1\over H\eta}=-{\sqrt{\ap} \over c_1\eta},
\nonumber\\
&&
\xi(\eta)=a e^{-\phi/2} \sim (-\eta)^{{c_2\over 2c_1}-1}
\sim (-\eta)^{0.136} .
\label{38}
\eea
The resulting slope of the string-branch of the spectrum,
\beq
k^{3/2}|\psi_k| \sim k^{1.36},
\label{39}
\eeq
is thus flatter than the slope of 
the low-energy dilatonic branch
($\sim k^{3/2}$, see \cite{6,20}), as anticipated in \cite{10,21}.  

The particular power (\ref{39}) obtained in this
simple example should not be taken, however, as a firm
prediction for the string-branch of the graviton spectrum. The
constant asymptotic values of $H$ and $\fp$, parametrizing in
phase space the fixed points of the $\sg$-model $\b$-functions
associated to the string effective action \cite{11}, may indeed
differ from those given in eq. (\ref{37}) when a more
``realistic" model of background (including dilaton potential,
quantum loop corrections) is considered. It may be interesting
to point out, in particular, that for $H=$const and $\fp\ra 0$ we
have $\xi\ra (-\eta)^{-1}$, and the spectrum tends to be flat,
i.e. scale-invariant. The fact that a flat perturbation
spectrum may emerge from a high-curvature string phase
provides an important counter-example to the expectation
that, in  string cosmology, the perturbation spectra grow in
general too fast \cite{20,22} to have observable effects at
large angular scales.

\section{CONCLUSION}
\label{IV}

In this paper we have perturbed the string effective action
around a spatially flat gravi-dilaton background, and we have
found the higher-derivative corrections, up to first order in
$\ap$, to the low-energy equation of tensor perturbations.

Applying the results to the pre-big bang cosmological models we
have shown that the high-curvature terms, 
which have a crucial 
influence on the evolution in time of the background, do not
affect in a qualitative way the evolution in time of the 
perturbations, both inside and outside the horizon (modulo a
constant shift of the final amplitude).

The low-energy perturbation equation thus remains valid for an 
order-of-magnitude estimate of the spectrum. This result,
however, is a direct consequence of having a background with a
high-curvature phase in which $H$ and $\fp$ stay frozen at the
string scale. In more general backgrounds the time evolution of
perturbations, and the final spectrum, may be significantly
affected by the high-curvature corrections.

The perturbative computations of this paper have been
explicitly performed in the string frame, and with an
appropriate representation of the background fields that
eliminates higher-than-second derivatives from the effective
equations. The results about the perturbation spectrum,
however, are frame-independent, and are expected to remain
invariant under arbitrary field redefinitions, even when such
redefinitions modify the explicit form of the action. To a
modification of the background corresponds indeed a
modification of the perturbation equation, in such a way that
the solution of the new perturbation equation, in the new
background,  should be the same as the solution of the old
equation in the old background (as explicitly checked 
in \cite{6} for the transformation from the string to 
the Einstein frame).

Finally, the results reported in this paper about  tensor
perturbations are expected to be qualitatively valid
also in the case of scalar perturbations. For modes crossing the
horizon during the string phase, in particular, the scalar
perturbation spectrum should be  determined by the
constant values of $H$ and $\fp$, and should tend to a
scale-invariant spectrum in the limit $\fp \ra 0$, as discussed
in Sec. \ref{III} for the graviton case. 

\acknowledgments
This work was supported in part by EC contract No.
ERBCHRX-CT94-0488. I am grateful to Gabriele Veneziano for
many useful discussions. Special thanks are due to Edward W.
Kolb for raising stimulating questions about string corrections
to the graviton spectrum, that motivated in part the work
presented in this paper.

\appendix

\section{SECOND ORDER PERTURBATION OF THE ACTION}
\label{A}

Consider the homogeneous and isotropic gravi-dilaton 
background, parametrized in $d=3$ by
\beq
ds^2=N^2(t)dt^2-a^2(t)dx_i^2 , ~~~~ \phi=\phi(t).
\label{a1}
\eeq
($i,j=1,2,3$). 
We define, as usual, $H=\dot a/a$ and $F= \dot N /N$. With the 
conventions 
$g_{\mu\nu}=(+---)$, 
$R_{\mu\nu\a}\,^\b=\pa_\mu\Ga_{\nu\a}\,^\b-...$, and 
$R_{\nu\a}=R_{\mu\nu\a}\,^\mu$, 
we compute the scalar curvature,
\beq
R={1\over N^2}\left(6HF-6\dot H-12 H^2\right),
\label{a2}
\eeq
and the Gauss-Bonnet invariant,
\bea
R_{GB}^2 &=& R_{\mu\nu\a\b}R^{\mu\nu\a\b}-4R_{\mu\nu}R^{\mu\nu}
+R^2 \nonumber\\
&=&
{24\over N^4}\left(\dot H H^2+H^4-FH^3 \right).
\label{a3}
\eea
Summing up all contributions, and setting $a=e^{\b}$, $H=\bp$, $ \dot H 
=\bpp$, the string effective action (\ref{21}) becomes, in this 
background,
\bea
&& S=\int dt e^{3\b-\phi}\Bigg[{1\over N} \left(6\bpp+12\bp^2-
6\bp F -\fp^2\right)\nonumber\\
&& \qquad 
+ {k\ap\over 4 N^3}\left(24\bpp \bp^2+24\bp^4-24\bp^3 F-\fp^4 
\right)\Bigg].
\label{a4}
\eea
The derivatives of the lapse function can be eliminated, by noting that 
\bea
&&{e^{3\b-\phi}\over N}\left(6\bpp-6\bp    F\right)=\nonumber\\
&&{e^{3\b-\phi}\over N}\left(6\bp\fp-18\bp^2\right)+
{d\over dt}\left(6\bp {e^{3\b-\phi}\over N} \right)
\label{a5}
\eea
and
\bea
&&{e^{3\b-\phi}\over N^3}\left(24\bpp \bp^2+24\bp^4   
-24\bp^3F\right)=\nonumber\\
 && 8\fp\bp^3{e^{3\b-\phi}\over N^3}+ {d\over dt}\left(8\bp
^3{e^{3\b-\phi}\over N^3} \right).
 \label{a6}
\eea
Using these results into eq. (\ref{a4}) we recover the action (\ref{23}),
modulo a total derivative that does not contribute to the equations of
motion. 

Let us now compute the second-order variation of the action (\ref{21}),
for a transverse and traceless metric 
perturbation $\da g_{\mu\nu}
=h_{\mu\nu}(x,t)$, with $h_\mu\,^\mu=0=\nabla_\nu h_\mu\,^\nu$ (the
indices of $h_{\mu\nu}$ are raised with the unperturbed metric
$g^{\mu\nu}$). We work in the synchronous gauge, where
\bea
&& 
g_{00}=1, ~~~g_{0i}=0, ~~~ g_{ij}=-a^2\da_{ij}, \nonumber \\
&&
h_{00}=0, ~~~h_{0i}=0, ~~~ g^{ij}h_{ij}=0, ~~~ \pa_j h_i\,^j =0 .
\label{a7}
\eea
The non-vanishing components of the Christoffel connection and of the
Ricci and Riemann tensor, for the unperturbed metric, are given by
\bea
&&
\Ga_{0i}\,^j=H\da_i^j , ~~~  \Ga_{ij}\,^0=-Hg_{ij}, \nonumber\\
&&
R_{00}=-3 (\dot H +H^2), ~~~ R_{ij}= -g_{ij}  (\dot H +3H^2), 
\nonumber \\
 &&
R_{0i0j}=g_{ij} (\dot H +H^2), ~~~ R_{ikjl}=
 H^2(g_{ij}g_{kl}-g_{kj}g_{il}).
\label{a8}
\eea
In this gauge, 
\bea
&&
\nabla_0 h_i\,^j=\hp_i\,^j, ~~\nabla_0h_{ij}=\hp_{ij}-2Hh_{ij},
~~\nabla_0\nabla_0 h_i\,^j=\hpp_i\,^j, \nonumber\\
&&
\nabla_0\nabla_0h_{ij}=\hpp_{ij}-2\dot H
h_{ij}-4H\hp_{ij}+4H^2h_{ij} 
\label{a9}
\eea
($\nabla_0$ is the covariant time derivative, while a dot denotes
partial time derivative). From these equations one easily derives the
identities (\ref{211}), useful to express all perturbed quantities in
powers of the convenient variable $h_i\,^j$ and of its derivatives. 

The non-vanishing controvariant components of the metric tensor, at
first and second order in $h$, are given by
\beq
\da^{(1)} g^{ij}=-h^{ij}, ~~~~ \da^{(2)} g^{ij}=h^{ik}h_k\,^j.
\label{a10}
\eeq
For the determinant of the metric tensor we thus obtain
\beq
\da^{(1)} \sqrt{-g}=0, ~~~~ \da^{(2)} \sqrt{-g}=-{1\over
4}a^3h_i\,^j h_j\,^i. 
\label{a11}
\eeq
The non-vanishing components of the perturbed connection are, at first
order, 
\bea
&&
\da^{(1)} \Ga_{0i}\,^j={1\over 2}\hp_i\,^{j}, ~~~~ 
\da^{(1)}\Ga_{ij}\,^0=-{1\over 2}\hp_{ij}, \nonumber \\ 
&&
\da^{(1)}\Ga_{ij}\,^k={1\over 2}\left(\pa_i h_j\,^k+
\pa_j h_i\,^k-\pa^k h_{ij}\right),
\label{a12}
\eea
and, at second order,
\bea
&&
\da^{(2)} \Ga_{0i}\,^j=-{1\over 2}\hp_i\,^{k}h_k\,^j, \nonumber \\ 
&&
\da^{(2)}\Ga_{ij}\,^k=-{1\over 2}h^k\,_l\left(\pa_i h_j\,^l+
\pa_j h_i\,^l-\pa^l h_{ij}\right). 
\label{a13}
\eea

At first order, and using the identities (\ref{211}), the non-vanishing
components of the Ricci tensor can be given in terms of  $h_i\,^j$ as 
\bea
&&
\da^{(1)}R_{i}\,^j=-{1\over 2}\left(\hpp_i\,^j
+3H \hp_i\,^j -{\nabla^2\over a^2} h_i\,^j\right)\equiv 
-{1\over 2} \Box h_i\,^j , \nonumber \\
&& \qquad
\da^{(1)}R_{ij}=-{1\over 2}g_{ik}\Bigg(\hpp_j\,^k+2\dot H h_j\,^k 
+3H \hp_j\,^k 
\nonumber \\
&&
\qquad
+6H^2 h_j\,^k -{\nabla^2\over a^2} h_j\,^k \Bigg)
\label{a14}
\eea
(where $\nabla^2=\da^{ij}\pa_i\pa_j$). At second order, 
\bea
&& 
\da^{(2)}R_{00}=\da^{(2)}R_{0}\,^0\nonumber\\
&&
={1\over 2}\left(\hpp_i\,^j h_j\,^i+{1\over 2}\hp_i\,^j \hp_j\,^i 
+2H \hp_i\,^j h_j\,^i \right), \nonumber\\
&&
\da^{(2)}R_{ij}={H\over 2} g_{ij}h_k\,^l\hp_l\,^k +
{1\over 4} g_{ik}\hp_j\,^l\hp_l\,^k \nonumber \\
&&
+{1\over 4} g_{jk}\hp_i\,^l\hp_l\,^k 
-{1\over 4}\pa_i h_k\,^l\pa_j h_l\,^k+
{1\over 2}\pa_kh_i\,^l\pa^k h_{lj} , \nonumber\\
&&
\qquad
\da^{(2)}R_{i}\,^j={1\over 2}\Bigg(h_i\,^k\hpp_k\,^j
+3H h_i\,^k\hp_k\,^j \nonumber\\
&&
\qquad
+H\da_i^jh_k\,^l\hp_l\,^k 
+\hp_i\,^k \hp_k\,^j -{1\over 2}\pa_i h_l\,^k \pa^j h_k\,^l \Bigg)
\label{a15}
\eea
(in the variation of $R_{ij}$ we have neglected all terms that after
integration by parts do not contribute to the perturbed action, because
of the gauge condition $\pa_i h_j\,^i=0$). 

We can now compute the second order perturbation of the various
terms appearing in the string effective action. We adopt, for simplicity,
a matrix notation for $h_i\,^j$,  setting $h_i\,^j h_j\,^i= {\rm Tr}~
h^2$,   $h_i\,^j \hp_j\,^i= {\rm Tr}~ (h \hp)$, and so on. From eqs.
(\ref{a10})-(\ref{a15}) we obtain 
\beq
\da^{(2)}\left(\sqrt{-g}\pa_\mu\phi\pa^\mu \phi\right)=
-{1\over 4}a^3\fp^2  {\rm Tr}~ h^2 ,
\label{a16}
\eeq
\beq
\da^{(2)}\left[\sqrt{-g}\left(\pa_\mu\phi\pa^\mu \phi\right)^2\right]=
-{1\over 4}a^3\fp^4  {\rm Tr}~ h^2 ,
\label{a17}
\eeq
\bea
&&
\da^{(2)}\left(\sqrt{-g} R\right)=R\da^{(2)}\sqrt{-g}+\nonumber \\
&&
+\sqrt{-g}\left(\da^{(1)}g^{\mu\nu}\da^{(1)}R_{\mu\nu}+
R_{\mu\nu}\da^{(2)}g^{\mu\nu}+
g^{\mu\nu}\da^{(2)}R_{\mu\nu}\right)\nonumber \\
&&
\qquad
=a^3  {\rm Tr}~\Bigg[h^2\left({3\over 2}\dot H +3H^2\right) + h \hpp +4H
h \hp\nonumber \\
&& \qquad
+{3\over 4} \hp^2 -{h\over 4}{\nabla^2\over a^2}h \Bigg],
\label{a18}
\eea
\bea
&&
\da^{(2)}\left(\sqrt{-g} R^2\right)=R^2\da^{(2)}\sqrt{-g}+\nonumber \\
&&
+2\sqrt{-g}R\left(\da^{(1)}g^{\mu\nu}\da^{(1)}R_{\mu\nu}+
R_{\mu\nu}\da^{(2)}g^{\mu\nu}+
g^{\mu\nu}\da^{(2)}R_{\mu\nu}\right)\nonumber \\
&&
\qquad
=-6a^3 (\dot H +2 H^2) {\rm Tr}~\Bigg[h^2\left({3\over 2}\dot H
+3H^2\right) \nonumber \\
&& \qquad
+ 2h \hpp +8H h \hp+{3\over 2} \hp^2 -{h\over 2}{\nabla^2\over a^2}h
\Bigg],
 \label{a19}
\eea
\bea
&&
\da^{(2)}\left(\sqrt{-g}
R_{\mu\nu}R^{\mu\nu}\right)=
(R_{\mu\nu})^2\da^{(2)}\sqrt{-g}\nonumber \\ 
&&
+\sqrt{-g}\left(\da^{(1)}R_\mu\,^\nu\da^{(1)}R_\nu\,^\mu+
2R_\mu\,^{\nu}\da^{(2)}R_\nu\,^\mu\right)\nonumber \\
&&
\qquad
=a^3{\rm Tr}~\Bigg[-h\hp (12 \dot H H +24 H^3)
-{\hp^2\over 2}(5\dot H +{18\over 4}H^2)\nonumber \\
 && \qquad
-h^2\left({3}{\dot H}^2+9\dot H H^2+9H^4\right)-h \hpp (4\dot H + 6H^2) \nonumber \\
 && \qquad
+{\hpp^2\over 4}+{3\over 2}H \hp \hpp + {1\over 4}\left({\nabla^2\over
a^2} h\right)^2 \nonumber \\
&& \qquad
-{1\over 2}(\hpp+3 H\hp-\dot H h -3 H^2 h){\nabla^2\over
a^2} h \Bigg].
\label{a20}
\eea

Finally, to complete the perturbation of the Gauss-Bonnet invariant,
we need the perturbations of the Riemann tensor. We find, to first
order,
\bea
 &&
\da^{(1)}R_{0i}\,^{0j}={1\over 2} \left(\hpp_i\,^j+2 H \hp_i\,^j\right) ,
\nonumber \\
 &&
\da^{(1)}R_{0i}\,^{jk}={1\over 2} \left(\pa^j\hp_i\,^k-
\pa^k\hp_i\,^j\right), \nonumber \\
 &&
\da^{(1)}R_{ik}\,^{0j}={1\over 2} \left(\pa_i\hp_k\,^j-
\pa_k\hp_i\,^j\right), \nonumber \\
 &&
\da^{(1)}R_{ik}\,^{jl}=\nonumber\\
&&
={1\over 2} \left(\pa_i\pa^j h_k\,^l-
\pa_k\pa^j h_i\,^l+\pa_k\pa^l h_i\,^j-\pa_i\pa^l h_k\,^j\right)
\nonumber\\
&&
+{H\over 2} \left(\da_i^j \hp_k\,^l-
\da_k^j \hp_i\,^l+\da_k^l \hp_i\,^j-\da_i^l \hp_k\,^j\right). 
\label{a21}
\eea
At second order, what we need for the perturbation of
$R_{\mu\nu\a\b}^2$ are the mixed components
\beq
\da^{(2)}R_{0i}\,^{0j}=-{1\over 2} \left(\hpp_i\,^kh_k\,^j +{1\over 2}
\hp_i\,^k\hp_k\,^j+2 H \hp_i\,^k h_k\,^j\right) ,
\label{a22}
\eeq
and the contraction
\bea
&&
R_{jl}\,^{ik}\da^{(2)}R_{ik}\,^{jl}=\nonumber\\
&&
=-{H^2\over 2} \left(\hp_i\,^j \hp_j\,^i + 8H h_i\,^j \hp_j\,^i-
h_i\,^j{\nabla^2\over a^2}h_j\,^i \right)
\label{a23}
\eea
(we have neglected terms that, after integration by parts, vanish
because of the trasversality condition). Therefore
\bea
&&
\da^{(2)}\left(\sqrt{-g}
R_{\mu\nu\a\b}R^{\mu\nu\a\b}\right)=
(R_{\mu\nu\a\b})^2\da^{(2)}\sqrt{-g}\nonumber \\ 
&&
+\sqrt{-g}\left(\da^{(1)}R_{\mu\nu}\,^{\a\b}\da^{(1)}
R_{\a\b}\,^{\mu\nu}+
2R_{\mu\nu}\,^{\a\b}\da^{(2)}R_{\a\b}\,^{\mu\nu}\right)\nonumber \\
&& \qquad
=(R_{\mu\nu\a\b})^2\da^{(2)}\sqrt{-g}+\sqrt{-g}\Bigg(
4\da^{(1)}R_{0i}\,^{0j}\da^{(1)}R_{0j}\,^{0i}\nonumber\\
&&\qquad
+2\da^{(1)}R_{0i}\,^{jk}\da^{(1)}R_{jk}\,^{0i}+2\da^{(1)}
R_{ik}\,^{0j}
\da^{(1)}R_{0j}\,^{ik}\nonumber\\
&&\qquad
+\da^{(1)}R_{ik}\,^{jl}\da^{(1)}R_{jl}\,^{ik}+8R_{0i}\,^{0j}
\da^{(2)}R_{0j}\,^{0i}\nonumber\\
&&\qquad
+2R_{jl}\,^{ik}\da^{(2)}R_{ik}\,^{jl} \Bigg)\nonumber\\
&&\qquad
=a^3 {\rm Tr}~\Bigg[\hp^2(2H^2-2\dot H)-h\hpp
(4H^2+4\dot H)\nonumber\\
&&\qquad
-h^2(3{\dot H}^2+6\dot H H^2+6H^4)-h \hp(8H\dot H +16H^3)
\nonumber\\
&&\qquad
+\hpp^2+4H\hp\hpp +\left({\nabla^2\over a^2} h\right)^2+
2\hp{\nabla^2\over a^2}\hp \nonumber\\
&&\qquad
+(H^2h -2H\hp){\nabla^2\over a^2}h \Bigg], 
\label{a24}
\eea
modulo a total derivative that does not contribute to the action.

Summing up the results (\ref{a16})--(\ref{a20}) and (\ref{a24}) we
finally obtain the action (\ref{212}) reported in Sec. \ref{II}.

\end{document}